# Thermalization of Starlight in the Steady-State Cosmology

Michael Ibison

*Institute for Advanced Studies at Austin*
*11855 Research Boulevard, Austin, TX 78759, USA*

September 2005

**Abstract.** We investigate the fate of starlight in the Steady-State Cosmology. We discover that it is largely unaffected by the presence of ions in intergalactic space as it gets progressively red-shifted from the visible all the way down to the plasma frequency of the intergalactic matter. At that point, after about 450 Gyr - and contrary to previously published claims - the radiation will be thermalized. Under the assumptions adopted by Gold, Bondi, Hoyle, Narlikar, Burbidge and others concerning the creation of matter in the Steady-State Cosmology, and using reasonable estimates for the baryonic mass-density and mass-fraction of $^4$He, the analysis predicts a universal radiation field matching the CMB, i.e. having a black-body spectrum and temperature of about 2.7 K. The Steady-state Cosmology predicts that this radiation field will appear to originate from the intergalactic plasma.

## INTRODUCTION

### Background

A commonly held view is that the discovery of the Cosmic Microwave Background (CMB) radiation by Penzias and Wilson in 1965 [1] delivered a fatal blow to the Steady-State Cosmology (SSC), whilst at the same time providing crucial evidence in favor of the Big Bang Cosmology. Weinberg [2] admits the difficulties, but nonetheless takes the time to ask how a nearly perfect universal black-body spectrum of EM radiation at low temperature might arise in the SSC as a by-product of matter-creation demanded by that theory. The conclusion is that such a process, though viable, looks artificial and is without the justification of necessity. However, proponents of the SSC, notably Bondi, Gold and Hoyle [3] had earlier predicted the presence of 'far-infrared' radiation on the basis of the following calculation. They assumed that matter creation in the SSC is initially prominently in the form of Hydrogen atoms or protons and electrons (the difference is not significant here). It followed that the presence, most prominently of $^4$He, must be accompanied by radiation the quantity of which could then be predicted from the relative abundances of Hydrogen and the energetics of the reaction. The average universal density of galactic matter was estimated at that time to be $3\times10^{-31}$ g/cm$^3$. The Helium mass-fraction of the visible baryonic density ($Y$) is about 25%. Given that 1 gm of H converted to $^4$He yields produces $6\times10^{11}$ erg, one expects a universal radiation field having energy density $4.5\times10^{-13}$ erg/cm$^3$. Bondi et al had suggested that this radiation might be found in the far-infrared, but had not associated it with a temperature. The reason, reported in the book by Hoyle, Burbidge and Narlikar [4], is that there was some disagreement between the authors about the likelihood that the radiation would be thermalized, in particular because no thermalizing mechanism could be identified at the time. As has subsequently been pointed out, $4.5 \times 10^{-13}$ erg/cm$^3$ is equal to the energy density of blackbody radiation at 2.78 K. Hoyle et al [4] argue persuasively that the history of Cosmology would be quite different if Bondi et al published this temperature in 1955, rather than just the energy density. Note that this energy density exceeds by a factor of about 45 the energy density of visible starlight, which is about $10^{-14}$ erg/cm$^3$. From the perspective of the SSC this is a secondary issue; some additional mechanism must be operating outside of stars to convert the Hydrogen to Helium in accord with the observed mass density ratio, as argued by Burbidge [5]. Accordingly the subject of this document is not exclusively starlight, but all radiation presumed to have been generated by nucleosynthesis within the Steady-State Cosmology.

Since then, proponents of the SSC, and more recently the Quasi-SSC, have focused on the role of carbon and metallic filaments - 'whiskers' - as a medium for absorption and subsequent thermalization of starlight [6]. The conclusion reported here does not detract from that work. It says however, that even if no such mechanism existed, thermalization of starlight is not just possible but inevitable - that the CMB has a natural explanation in the Steady-State-Cosmology.

## THE INTERGALACTIC PLASMA

### Propagation Constant

Here and throughout this document we are concerned with the fate of a typical Fourier component of radiation after it has left a star, as it travels through largely empty space, populated by highly rarified ionized matter – mostly electrons and protons. A single such component at radian frequency $\omega$ of a component $A_\mu$ of the 4-potential propagating in the *x* direction though a macroscopic medium can be written

$$A_\mu(t,x) = A_\mu(0,0)\exp(in\omega(t-x)). \tag{1}$$

(Here and throughout we use Gaussian units and $c=1$). *n* is the (possibly complex) refractive index accounting for the dielectric and dissipative properties of the intergalactic plasma. Davies [7,8] gives the following expression for $n^2$ - the complex relative permittivity of the plasma:

$$\varepsilon_r = n^2 \equiv (n_0 - i\kappa)^2 = \left\{1 - \frac{\omega_p^2/\omega^2}{1+\left[\omega\tau_e + \frac{\omega_{coll}}{\omega}\right]^2}\right\} - i\left\{\frac{\omega_p^2/\omega^2\left[\omega\tau_e + \frac{\omega_{coll}}{\omega}\right]}{1+\left[\omega\tau_e + \frac{\omega_{coll}}{\omega}\right]^2}\right\} \tag{2}$$

which can also be written

$$\varepsilon_r = 1 - \frac{\omega_p^2}{\omega^2\left(1 - i\left(\omega\tau_e + \frac{\omega_{coll}}{\omega}\right)\right)}. \tag{3}$$

Since here we are interested in particular in the damping of radiation as it propagates through the intergalactic plasma, we need focus only on the imaginary part of the refractive index. More specifically, for as long as the WKB approximation holds, a typical Fourier component of radiation from a 'point' source will suffer, in addition to the usual $1/r^2$ spatial fall-off in intensity $I$, an exponential decay rate which one can characterize by an exponent $\alpha(t)$:

$$I(t) \sim \frac{I_0}{r(t)^2} e^{-\alpha(t)}. \tag{4}$$

Here $r(t)$ is the coordinate distance - i.e. satisfying $dt = a(t)dr(t)$ in the case of the flat-space FRW metric - traveled by a typical photon at the frequency in question. If we take the time of emission as $t = 0$ then the exponential decay rate is

$$\alpha(t) = 2\int_0^t dt\,\text{Im}(n(\omega))\omega \tag{5}$$

where $\omega$ is a function of time due to its cosmological red-shift. A necessary condition for thermalization is that $\alpha > 1$, and preferably much greater then one, though this is not sufficient (see below).

The aim of the following is to determine if the plasma thermalizes starlight, and, if so: when, and by what process. First we examine the terms in the expression (3) in detail.

## Definitions and magnitudes of the contributions

*Electron Relaxation Time*

In (2) $\tau_e$ is the electron relaxation time

$$\tau_e = \frac{2e^2}{3m_e} = 6.3\times 10^{-24}\,s \tag{6}$$

which is the order of the time it takes light to travel the distance of the classical electron radius. It is evident from (2) and (3) that the term $\omega\tau_e + \omega_{coll}/\omega$ controls the damping in the propagation constant. However, the two contributions have quite different physical significance. The first is the due to Thomson scattering, which is the long wavelength $(\omega << m_e/\hbar)$ limit of Compton scattering. In this limit the scattering is entirely elastic. It appears in the dissipative part of the propagation constant associated with a single Fourier component because it takes energy away from the primary wave, scattering into secondary waves at the same frequency, but with changed momentum direction. Since this term does not represent true absorption its effects must somehow be subtracted out if one is attempting to associate decay of the Fourier component with absorption and therefore thermalization.

*Plasma Frequency*

In (2)

$$\omega_p \equiv \sqrt{\frac{4\pi n_i e^2}{m_e}} \tag{7}$$

is the plasma frequency of the medium where $n_i$ is the ion charge density and $m_e$ is the mass of the charge electron. We compute the ion number density as follows: Assume that the majority of baryonic matter is contained in the intergalactic medium and is ionized. Assume that the baryonic matter density is 25% of the critical density, the latter being given by

$$\rho_{crit} = \frac{3H_0^2}{8\pi G} \tag{8}$$

where $H_0$ is the Hubble constant. Then the ion number density is

$$n_i = 0.25 \frac{3H_0^2}{8\pi G m_p}. \tag{9}$$

Using the current estimate for the Hubble parameter (PDG 2004)

$$h \equiv 9.78 \text{ Gyr } H_0 = 0.73 \tag{10}$$

which therefore corresponds to a Hubble time of $1/H_0 = 13.4 \, Gyr$, (9) gives that the ion number density is $n_i \approx 1.5/m^3$. Putting this in (7) gives a plasma frequency of 70 rad/s, i.e. $f_p \equiv \omega_p / 2\pi = 11 \text{ Hz}$.

*Collision Frequency*

In (2) $\nu_{coll}$ is the mean collision frequency between electrons and ions in the plasma, and is a function of the plasma temperature. When the EM photon energy is much less than the thermal energy of the plasma Ginzburg [9] gives

$$\lim_{\omega \to 0} (\omega_{coll}) \equiv \omega_{coll}(0) = \frac{\pi e^4}{(k_B T_i)^2} \bar{v}_e n_i \tag{11}$$

where $T_i$ is the temperature of the ions, and $\bar{v}_e$ is the mean speed of the electrons. Assuming that the ions and the electrons are in equilibrium, the electrons will follow a Maxwell velocity distribution with temperature $T_i$. Their mean speed is therefore [10]

$$\bar{v}_e = \sqrt{\frac{8 k_B T_i}{\pi m_e}} \tag{12}$$

and so the collision frequency is given by

$$\omega_{coll}(0) = \frac{e^4 n_i}{(k_B T_i)^{3/2}} \sqrt{\frac{8\pi}{m_e}}. \tag{13}$$

Let us use the definition of the plasma frequency (7) to replace the ion density in (13):

$$n_i = \frac{m_e \omega_p^2}{4\pi e^2} \Rightarrow \omega_{coll}(0) = \frac{e^2 \omega_p^2}{(k_B T_i)^{3/2}} \sqrt{\frac{m_e}{2\pi}}. \qquad (14)$$

An expression valid at arbitrary EM radiation frequencies involves, according to Davies [7,8], a factor of the form $1 - \exp(-\hbar\omega/k_B T_i)$. It follows that the more general result which reduces to the form (13) when $\hbar\omega \ll k_B T_i$ must be

$$\omega_{coll} = \frac{k_B T_i}{\hbar\omega}\left(1 - \exp\left(-\frac{\hbar\omega}{k_B T_i}\right)\right) \omega_{coll}(0). \qquad (15)$$

(Eq. (5.93) in [7] contains an error – the factor $m_e/4\pi e^2$ should read $(m_e/4\pi e^2)^{-1}$.) With (15), the more general expression for the collision frequency (13) is

$$\omega_{coll} = \frac{e^2 \omega_p^2}{\hbar\omega}\sqrt{\frac{m_e}{2\pi k_B T_i}}\left(1 - \exp\left(-\frac{\hbar\omega}{k_B T_i}\right)\right). \qquad (16)$$

Let us now assume that the plasma temperature is that of the CMB i.e. $T_i = T_{CMB} = 2.7\text{K}$ and define a frequency

$$f_{CMB} \equiv \frac{k_B T_{CMB}}{h} = 56 \text{ GHz} \qquad (17)$$

(this is exactly 1/3 the frequency of the spectrum maximum of a Planck distribution at this temperature). Then (16) can be written

$$\omega_{coll} = \frac{e^2 \omega_p^2}{\hbar\omega}\sqrt{\frac{m_e}{2\pi\hbar\omega_{CMB}}}\left(1 - e^{-\omega/\omega_{CMB}}\right). \qquad (18)$$

where $\omega_{CMB} \equiv 2\pi f_{CMB}$. Further simplification is possible if we introduce the Compton frequency of the electron $\omega_c$, and the fine structure constant $\alpha = e^2/\hbar$. With these, the collision frequency can be written

$$\omega_{coll} = \frac{\alpha}{\sqrt{2\pi}}\sqrt{\frac{\omega_c}{\omega_{CMB}}}\frac{\omega_p^2}{\omega}\left(1 - e^{-\omega/\omega_{CMB}}\right). \qquad (19)$$

Clearly there are two regimes:

$$\omega_{coll} = \begin{cases} \dfrac{\alpha}{\sqrt{2\pi}}\sqrt{\dfrac{\omega_c}{\omega_{CMB}}}\dfrac{\omega_p^2}{\omega}; & \omega \gg \omega_{CMB} \\ \dfrac{\alpha}{\sqrt{2\pi}}\sqrt{\dfrac{\omega_c}{\omega_{CMB}}}\dfrac{\omega_p^2}{\omega_{CMB}} & \omega \ll \omega_{CMB} \end{cases}. \qquad (20)$$

Putting in the numbers one finds that the low frequency limit is

$$f_{coll}(0) \equiv \frac{\omega_{coll}(0)}{2\pi} = 2.9 \times 10^{-7} \text{ Hz}. \qquad (21)$$

That is, the mean time between collisions in the low frequency limit of radiation is $3.4 \times 10^6$ s $\approx$ 40 days. A plot of $f_{coll}(f)$ is given in Figure 1.

Connection can be made with the result given by Davies (Eq. (5.93) in [7], and corrected as indicated above):

$$\omega_{coll} = \frac{m_e n_i}{4\pi e^2 \sqrt{T_i} \omega} A \left(1 - \exp\left(-\frac{\hbar\omega}{k_B T_i}\right)\right) \quad (22)$$

where we are told that $A$ is factor depending on $k_B$, $e$, and $m_e$ (Davies chooses units in which $\hbar = 1$). Comparing (22) with (16) and using (7) it is inferred that

$$A = \frac{4e^6}{\hbar\sqrt{k_B}} \left(\frac{2\pi}{m_e}\right)^{3/2}. \quad (23)$$

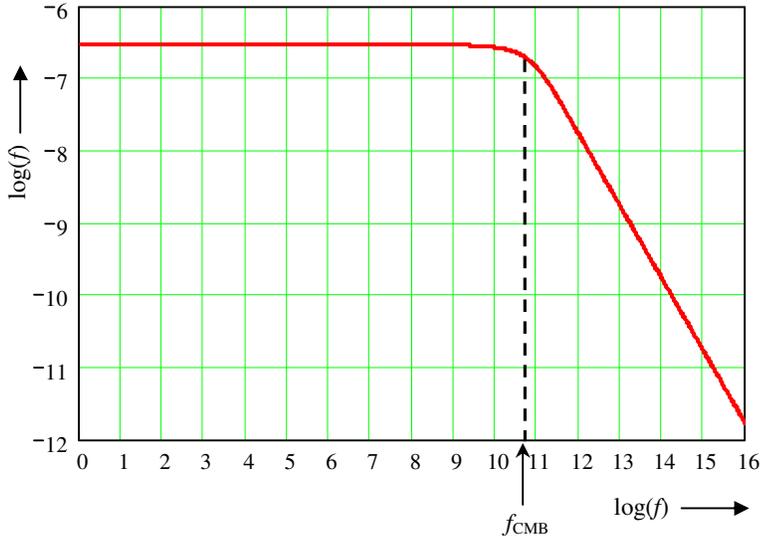

**FIGURE 1.** Plot of the function $f_{coll}(f)$ defined implicitly in (18).

## PROPAGATION

### Cosmological Red-shift

The traditional (Friedmann Robertson Walker) form of the cosmological line element in flat space is

$$ds^2 = dt^2 - a^2(t)d\mathbf{r}^2 \quad (24)$$

where, in the steady-state cosmology,

$$a(t) = e^{H_0 t}, \quad (25)$$

where we have taken the present era ($a_0 = 1$) to be at $t = 0$. With this, the cosmological redshift of a typical Fourier component obeys

$$\omega(t) = \omega_0 / a(t) = \omega_0 e^{-H_0 t} \quad (26)$$

where $\omega_0$ is the radian frequency at the time of emission, taken here to be at $t=0$. With this, the integral over time in (5) can be written

$$\int_0^{t_f} dt = \int_{\omega_0}^{\omega_f} d\omega \left(\frac{d\omega}{dt}\right)^{-1} = \int_{\omega_f}^{\omega_0} d\omega \frac{1}{H_0 \omega} \tag{27}$$

where $t_f$ is the time at which the radiation is eventually absorbed, and $\omega_f$ is the radian frequency at that time. With this the absorption defined in (5) can be written

$$\alpha(t_f) = \frac{2}{H_0} \int_{\omega_f}^{\omega_0} d\omega \, \text{Im}(n(\omega)). \tag{28}$$

## Effect of Collisions

As discussed under the subheading of electron relaxation time, collisions - though not Thomson scattering - represent a genuine absorptive process by which the radiation may be brought into its maximum entropy state. There is no straightforward way to separate these two contributions to the permittivity because they do not appear additively either in the permittivity or, more importantly, in the imaginary part of the refractive index (i.e. the imaginary part of the square root of the expression (2)). However, at the very least, one can be sure that if one sets the Thomson scattering to zero in the original expression, what remains will be a minimum estimate for the component of absorptive damping, $\alpha_{abs}$ say, in the expression (5)[1]. That is, we assert that

$$\alpha_{abs}(t_f) \geq \alpha_{\min}(t_f) \equiv 2 \int_0^{t_f} dt \, \omega \, \text{Im}\left(n(\omega)\big|_{\tau_e=0}\right) \tag{29}$$

which, using (28), gives

$$\alpha_{\min}(t_f) = \frac{2}{H_0} \int_{\omega_f}^{\omega_0} d\omega \, \text{Im}\left(n(\omega)\big|_{\tau_e=0}\right). \tag{30}$$

From (3) one has

$$n(\omega)\big|_{\tau_e=0} = \sqrt{1 - \frac{(\omega_p/\omega)^2}{1 - i\omega_{coll}/\omega}}. \tag{31}$$

A typical component of starlight, having wavelength 500 nm say, has frequency $\omega_0/2\pi = 6\times 10^{14}$ Hz - which, of course, is greater than the plasma frequency by many orders of magnitude. Given this, it is clear from (31) that the refractive index will remain close to unity as the red-shifted starlight falls below the collision frequency. More particularly the magnitude of the term $(\omega_p/\omega)^2/(1-i\omega_{coll}/\omega)$ remains small compared to unity for all frequencies well above the plasma frequency. Initially therefore, let us proceed assuming that absorption takes place before that point, and require that $\omega_f \gg \omega_p$ on the right hand side of (30), with the aim of testing that assumption for self-consistency through satisfaction or otherwise of the inequality. Then (31) can be expanded binomially, and

$$\alpha_{\min}(t_f) = \frac{1}{H_0} \int_{\omega_f}^{\omega_0} d\omega \, \text{Im}\left(\frac{(\omega_p/\omega)^2}{1 - i\omega_{coll}/\omega}\right) = \frac{\omega_p^2}{H_0} \int_{\omega_f}^{\omega_0} d\omega \frac{\omega_{coll}}{\omega^3 \left(1 + (\omega_{coll}/\omega)^2\right)}. \quad (32)$$

Where, recalling (19), $\omega_{coll}$ is a function of the frequency.

*Assumption that absorption occurs before starlight falls below microwave background*

Let us first estimate the integral in the case that $\omega_f \gg \omega_{CMB}$ i.e. supposing that the absorption is complete before the starlight is red-shifted to the microwave frequency, testing for self-consistency. If so, then one can use the second of (20), whence (32) becomes

$$\alpha_{\min}(t_f) = \alpha \sqrt{\frac{\omega_c}{2\pi\omega_{CMB}}} \frac{\omega_p^4}{H_0} \int_{\omega_f}^{\omega_0} d\omega \frac{1}{\omega^4 + \frac{\alpha^2}{2\pi} \frac{\omega_c \omega_p^4}{\omega_{CMB}}}. \quad (33)$$

It is convenient to normalize all frequencies with respect to $\omega_{CMB}$, which will be denoted by a tilde,

$$\alpha_{\min}(t_f) = \alpha \sqrt{\frac{\tilde{\omega}_c}{2\pi}} \frac{\omega_{CMB} \tilde{\omega}_p^4}{H_0} \int_{\tilde{\omega}_f}^{\tilde{\omega}_0} d\tilde{\omega} \frac{1}{\tilde{\omega}^4 + \frac{\alpha^2}{2\pi} \tilde{\omega}_c \tilde{\omega}_p^4}. \quad (34)$$

where now the condition for self-consistency is that $\alpha_{\min}(t_f) = 1$ can be achieved with some $\tilde{\omega}_f \gg 1$. Using the numbers above one finds that $\alpha^2 \tilde{\omega}_c \tilde{\omega}_p^4 / 2\pi = 2.7 \times 10^{-35}$. Given this and that $\tilde{\omega} \gg 1$ throughout the range of integration one can safely ignore this term in the denominator. Performing the integration one then has

$$\alpha_{\min}(t_f) = \frac{\alpha}{3} \sqrt{\frac{\tilde{\omega}_c}{2\pi}} \frac{\omega_{CMB} \tilde{\omega}_p^4}{H_0} \left(\frac{1}{\tilde{\omega}_f^3} - \frac{1}{\tilde{\omega}_0^3}\right). \quad (35)$$

One finds for the second of these two terms that $\sqrt{\tilde{\omega}_c / 2\pi} \left(\alpha \omega_{CMB} \tilde{\omega}_p^4 / 3 H_0 \tilde{\omega}_0^3\right) = 8.15 \times 10^{-21}$ and is therefore negligible compared with 1. Solving for the value of $\tilde{\omega}_f$ to give the minimum acceptable criterion for absorption by collisions:

$$\tilde{\omega}_f = \left(\frac{\alpha}{3} \sqrt{\frac{\tilde{\omega}_c}{2\pi}} \frac{\omega_{CMB} \tilde{\omega}_p^4}{H_0}\right)^{1/3} = 2.15 \times 10^{-3}. \quad (36)$$

This result contradicts the assumption used to justify use of (33) in place of (32). It is concluded that collisions have a negligible effect on starlight by the time it has red-shifted to the frequency of the microwave background.

---

[1] Though the two components do not contribute additively to $\varepsilon_r$ - and therefore to the imaginary part of $n$ – it is important for the validity of this argument that they are always both of the same sign.

## Numerical evaluation

Eq. (32) with the general expression (19) for the collision frequency is

$$\alpha_{\min}(t_f) = \frac{\alpha \omega_p^4}{H_0} \sqrt{\frac{\omega_c}{2\pi\omega_{CMB}}} \int_{\omega_f}^{\omega_0} d\omega \frac{1 - e^{-\omega/\omega_{CMB}}}{\omega^4 \left(1 + \frac{\alpha^2 \omega_c \omega_p^2}{2\pi\omega_{CMB}\omega^2}\left(1 - e^{-\omega/\omega_{CMB}}\right)^2\right)}. \qquad (37)$$

Unfortunately the integral cannot be evaluated in terms of a finite series of standard functions. Instead, the expression on the right has been integrated numerically for variable $\omega_f$, the results of which are shown in Figure 2. The effect of the two regimes either side of the microwave frequency is clearly apparent. However, one sees that the absorption never approaches unity.

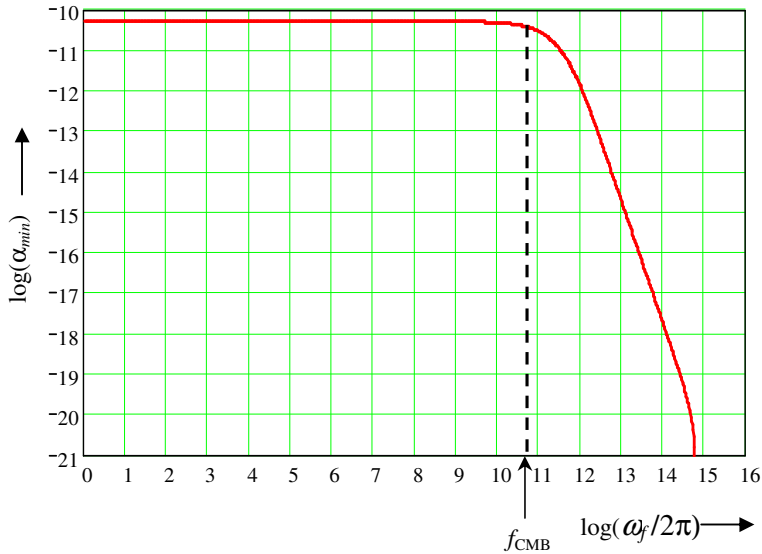

**FIGURE 2.** Plot of the minimum estimate for the absorption $\alpha_{abs}$ as defined in (37).

Bearing in mind that this analysis is valid only for red-shifted starlight remaining above the plasma frequency, it is concluded that the spectrum of visible starlight is nearly unchanged on its journey through the intergalactic plasma whilst undergoing red-shift from the visible down to the plasma frequency. That is, collisions play no role at all in the fate of the starlight. The same conclusion was reached by Davies [7,8], though those works employ a more qualitative argument based on a rougher approximation.

## Radiation Close To The Plasma Frequency

Eventually the starlight will be red-shifted to the plasma frequency of the intergalactic medium. Using the numbers above and the relation (26) one can compute the age of the arriving starlight:

$$\omega_p = \omega_0 e^{-H_0 t} \Rightarrow t = \frac{1}{H_0} \log \frac{\omega_0}{\omega_p} = 424 \text{ Gyr} . \tag{38}$$

At frequencies approaching the plasma frequency, $\omega \tau_e + \omega_{coll}/\omega$ remains small compared to unity. It follows that the relative permittivity given by (2) has a negative real part and an imaginary part whose magnitude is relatively small. Consequently the refractive index – the square root of the relative permittivity – will relatively quickly acquire a significant imaginary component. At this point the plasma looks like a conductor; the EM field variations are now sufficiently slow that the plasma electrons move to effectively neutralize the field. We will now discuss what happens next.

*Historical Misconception*

Generally, when an EM field strikes a conducting surface it is reflected. That is, when an EM field propagates from a dielectric region (where the frequency of radiation is greater than the plasma frequency) to a conducting region, (where the frequency of radiation is less than the plasma frequency), it is reflected at the interface between the two regions. With this in mind, Davies [7,8] predicted that the red-shifted starlight would reflect off the intergalactic plasma once it reached the plasma frequency. With this and the result of the analysis of the effect of collisions, he concluded that the intergalactic plasma plays no role at all in the fate of the starlight. This view was subsequently repeated by Pegg [11]. Davies went on to state that despite the ineffectiveness of the plasma in the steady state cosmology, discrete objects (e.g. stars) will nonetheless eventually absorb and thermalize the starlight. This conclusion was arrived at presumably because it was seen that the starlight cannot go on forever in a universe of matter with a constant coordinate density – it must eventually be thermalized by something. The argument leading to this conclusion however contradicts the above argument applied to the intergalactic plasma. If the latter fails to thermalize starlight, then it is not at all clear why agglomerations of matter would be more successful.

*Boundary conditions for radiation arriving at the plasma frequency*

The above reasoning incorrectly applies the analogy of the conducting reflector to the cosmological situation. One way to recognize the problem is to note that reflection, if it did occur, would presumably entail further propagation and continued redshift, taking the erstwhile starlight yet further below the plasma frequency. If so, the surrounding plasma would still look like a conductor – yet more so. That is, the radiation cannot escape the conductor by *spatial* reflection.

Returning to the analogy of reflection at a conductor - a plane mirror say at $z = L$ - the boundary between dielectric and conductor is a surface present for all time. As a result of the interaction at the boundary there is near perfect inversion of momentum and near zero loss of energy in the radiation. Correspondingly there is near zero transfer of energy to the mirror, which nonetheless acquires momentum nearly twice that of the incident radiation. The field may penetrate the surface to some degree, decaying exponentially into the medium. If the medium is lossless – a superconductor for example - then the decay does not represent absorption.

This analogy can be related to the cosmological case under consideration by exchanging the roles of space and time, noting that the conducting 'surface' now resides at a *time* $t = t_f$ say (the time the starlight arrives at the plasma frequency), present for all *space*. Taking advantage of the invariance in 1+1 D of the Maxwell equations under this exchange, it is deduced that as a result of the interaction at the boundary there is near perfect inversion of *energy* and near zero loss of *momentum* in the radiation. Correspondingly there is near zero transfer of *momentum* to the medium, which nonetheless acquires *energy* nearly twice that of the incident radiation.

It is inferred that Davies presumably employed the above picture of the plane mirror, applying it literally to the Cosmological case, rather than as an analogy with the space and time coordinates interchanged.

There are several caveats. The above suggests that the radiation is time-reversed at the plasma boundary. In this 1+1 D analogy there is no diffraction and therefore no signature of advanced and retarded propagation in the convergence or divergence of the radiation. Time-reversed radiation *is* expected to blue shift, but it does so as it propagates *backwards* in time. However, in the absence of diffraction, such radiation will be indistinguishable from the incident radiation that is red-shifting as it travels forwards in time. Consequently such radiation can always be reinterpreted – at least in 1+1 D - as having come from another source, symmetrically opposite that of the 'incident' that was the initial subject of the investigation. Similarly, the phrase 'negative energy' requires some interpretation. A 'reflected' wave having negative energy and traveling backwards in time can be reinterpreted as having positive energy and traveling forwards in time.

In practice the refractive index is not discontinuous and therefore the boundary condition does not occur abruptly as implied above. A more thorough treatment can be achieved by solving the macroscopic Maxwell system of equations in the curved space-time of (24). The intergalactic medium will then manifest as a time-varying complex permittivity. The fate of an initial excitation pulse at optical frequency can then be determined as the red-shifting pulse approaches and passes through the plasma frequency. At that point one still expects to see the pulse decay into the future in some manner not qualitatively different from the exponential decay at a hard interface.

## Thermalization

Taking these points into account, the important conclusion is that the boundary condition dictates that the energy of the radiation (from the pair of sources identified above) is nearly perfectly transferred to the medium, which here is the intergalactic plasma. That is, the radiation is *absorbed*. In the case of spatial reflection at a plane mirror, the idealized EM theory by itself does not tell us about the fate of the momentum transferred to the mirror. Similarly this qualitative analysis does not say what happens to the energy of the radiation absorbed by the plasma. It tells us that the energy is no longer in the EM field, but now resides in the predominantly mechanical degrees of freedom of the matter, which, in this case must be the kinetic energy of the electrons. Having shown some means of entropy-increasing transfer of energy from the radiation field to matter, it is simply a matter of time before equilibrium is attained, and since time is not in short supply in the steady state cosmology, it is concluded that equilibrium is inevitable.

This mechanism of thermalization could, in principle, be achieved in the laboratory, though it would not be without technical difficulties. Consider a short pulse of nearly monochromatic radiation propagating in a non-dissipative dielectric medium. Supposing the requisite technology exists, let the dielectric properties of the medium be changed en mass from dielectric to conductor (i.e. conducting at the frequency of the radiation in question) instantaneously - without appreciable propagation delay compared to that of light. Having nowhere to go (in space), the radiation is now effectively trapped inside a conductor, whose electrons will neutralize the electric field in the order of one period. In doing so they necessarily acquire the energy of the EM field, which must then be dissipated as heat by the mechanism of collisions of various kinds in the medium. The important point to notice is the crucial difference from conventional boundary conditions on the radiation: here the boundary is time-like rather than space-like. Though technically it would be difficult to achieve, there is no problem with the principle.

The role of the intergalactic plasma in the steady-state cosmology can be summarized as follows. Locally, at any typical location, the plasma absorbs energy from starlight that has been red-shifted to around 11 Hz, and which is typically around 450 Gyr old. As a result of collisions the plasma re-radiates the acquired energy as a secondary spectrum which must be self-reproducing upon further absorption and re-emission. That is, it must be the black-body equilibrium spectrum having the same density as the incident starlight, having, therefore, a temperature of about 2.7 K. Note that the secondary emission will generally be a frequency up-conversion (from 11 Hz to about 180 GHz). This is nonetheless in the direction of increasing total entropy; matter at 0 K and a radiation spectrum centered on 11 Hz - even if black-body and therefore having a temperature of around $1.6 \times 10^{-10}$ K – will have higher entropy than that of matter with a Maxwell-Boltzmann velocity distribution plus black-body radiation, both at 2.7 K.

## DISCUSSION

The conclusion of this work is that thermalization of starlight is inevitable in the Steady-State-Cosmology. In particular, if starlight is not otherwise thermalized beforehand, once it falls below the plasma frequency of the intergalactic plasma it will be thermalized there, approximately 450 Gyr after emission, for typical optical frequencies. Had this result been known (in 1955) it might have significantly influenced the historical development of Cosmology.

No attempt has been made here to examine the consequences of such a mechanism for the power spectrum of angular-fluctuations. It is not know if they will agree with the recent COBE and WMAP data any better than the models founded upon the Big Bang Cosmology and in particular the state of affairs towards the end of the radiation-dominated era. Even so, it seems likely that 'local' thermalization in the plasma of empty space might favor a closer approximation to perfect isotropy than models requiring density fluctuations in the hot plasma at the end of recombination, in accord perhaps with recent observations.

A prediction of the steady-state Cosmology is that the CMB temperature does not change. This appears to be in conflict with recent observations; the historical temperature has been inferred from the absorption spectrum of gas illuminated by quasar radiation [12-14]. Arguably these results are not yet strong enough to completely rule out the possibility of a constant temperature background, though at the least they demand a plausible alternative explanation from the Steady-State Cosmology.

## ACKNOWLEDGMENTS

I am very grateful to Sergei Levshakov for alerting me to the work of LoSecco and Mathews and his own very relevant work. Thank you also to Alain Blanchard for the educative discussions on some of the basic issues. And thank you also to Koch Raymond for alerting me to the shortcomings of the implicit WKB assumption.

## REFERENCES

1. A. A. Penzias and R. W. Wilson, *Astrophys. J.* 142 (1965) 419.
2. S. Weinberg, *Gravitation and Cosmology*, John Wiley & Sons, Inc., New York 1972.
3. H. Bondi, T. Gold, and F. Hoyle, *Observatory* 75 (1955) 80.
4. F. Hoyle, G. Burbidge, and J. V. Narlikar, *A Different Approach to Cosmology*, Cambridge University Press, Cambridge 2000.
5. G. Burbidge, *Publications of the Astronmical Society of the Pacific* 70 (1958) 83.
6. F. Hoyle and C. Wickramasinghe, *Astrophys. & Space Sci.* 147 (1988) 248.
7. P. C. W. Davies, *The Physics of Time Asymmetry*, University of California Press, Berkeley, CA 1977.
8. P. C. W. Davies, *J. Phys. A* 5 (1972) 1722-1737.
9. V. L. Ginzburg, *The Propagation of Electromagnetic Waves in Plasmas*, Pergamon Press Ltd., Oxford 1970.
10. E. H. Kennard, *Kinetic theory of gases*, McGraw-Hill, New York 1938.
11. D. T. Pegg, *Rep. Prog. Phys.* 38 (1975) 1339-1383.
12. J. M. LoSecco, *Phys. Rev. D* 64 (2001) 123002-1-123002/8.
13. P. Molaro et al, *Astronomy & Astrophysics* 381 (2002) L64-L67.
14. S. A. Levshakov, P. Molaro, and S. D'Odorico, in: Eds.: S. Kato and T. Shiromizu *New Trends in Theoretical and Observational Cosmology*, Universal Academy Press, 2002. 1-4.